\documentclass[twocol]{ametsoc}

\journal{jcli}

\bibpunct{(}{)}{;}{a}{}{,}

\title{Understanding decreases in land relative humidity with global warming: conceptual model and GCM simulations}

\authors{Michael P. Byrne\correspondingauthor{Michael P. Byrne, 
ETH Z\"{u}rich, 
Sonneggstrasse 5, 
8092 Z\"{u}rich, 
Switzerland}}

\affiliation{ETH Z\"{u}rich, Z\"{u}rich, Switzerland}

\email{michael.byrne@erdw.ethz.ch}

\extraauthor{Paul A. O'Gorman}
\extraaffil{Massachusetts Institute of Technology, Cambridge, Massachusetts}

\abstract{Climate models simulate a strong land-ocean contrast in the response of near-surface relative humidity to global warming: relative humidity tends to increase slightly over oceans but decrease substantially over land. Surface energy balance arguments have been used to understand the response over ocean but are difficult to apply over more complex land surfaces. Here, a conceptual box model is introduced, involving moisture transport between the land and ocean boundary layers and evapotranspiration, to investigate the decreases in land relative humidity as the climate warms. The box model is applied to idealized and full-complexity (CMIP5) general circulation model simulations, and it is found to capture many of the features of the simulated changes in land relative humidity. The box model suggests there is a strong link between fractional changes in specific humidity over land and ocean, and the greater warming over land than ocean then implies a decrease in land relative humidity. Evapotranspiration is of secondary importance for the increase in specific humidity over land, but it matters more for the decrease in relative humidity. Further analysis shows there is a strong feedback between changes in surface-air temperature and relative humidity, and this can amplify the influence on relative humidity of factors such as stomatal conductance and soil moisture.}

\begin{document}

\maketitle

%








\section{Introduction}



\begin{figure}[t]
  \centering
\noindent\includegraphics[width=19pc,angle=0]{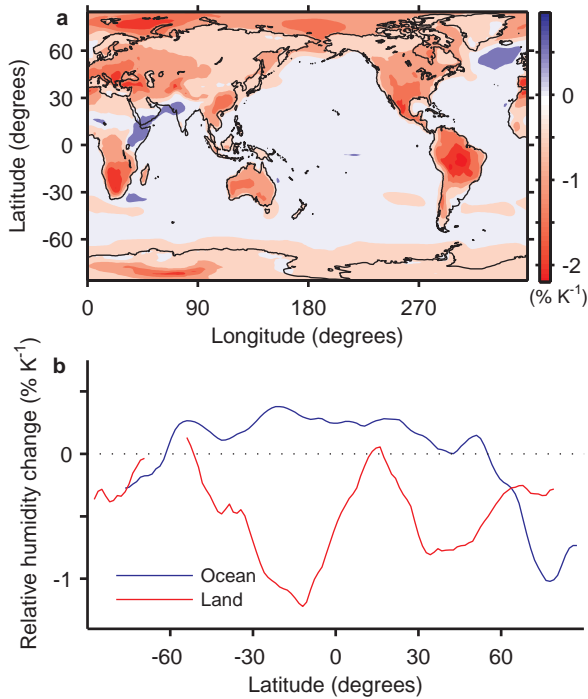}
  \caption{Multimodel-mean changes in surface-air relative humidity between the historical (1976-2005) and RCP8.5 (2070-2099) CMIP5 simulations, normalized by the global- and multimodel-mean surface-air temperature changes [(a) and (b)]. For (b), the zonal averages over all ocean (blue) and land (red) gridpoints are shown at each latitude. 
}
\label{fig:delta_hurs_CMIP5}
\end{figure}

Observations and climate-model simulations show a pronounced land-ocean warming contrast in response to a positive radiative forcing, with land temperatures increasing more than ocean temperatures \citep{manabe_1991, sutton_2007, byrne_ogorman_2013}. A land-ocean contrast is also found for the response of near-surface relative humidity in climate-model simulations, with small increases in relative humidity over ocean and larger decreases in relative humidity over continents \citep{ogorman_muller_2010,laine_et_al_2014,fu_feng_2014}. This land-ocean contrast in changes in relative humidity is clearly evident in Fig. \ref{fig:delta_hurs_CMIP5} for simulations from the Coupled Model Intercomparison Project 5 (CMIP5) that will be discussed in detail in sections 3 and 4. However, the long-term observational trends in near-surface relative humidity are not yet clear. Based on observations over 1975-2005, \citet{dai_2006} found a decreasing trend in surface relative humidity over ocean, but no significant trend over land. Later studies have found a sharp decrease in land relative humidity since 2000 \citep{simmons_et_al_2010, willett_et_al_2014}, and this is more consistent with the long-term climate-model projections.

Changes in land relative humidity are important for the land-ocean warming contrast \citep{byrne_ogorman_2013,byrne_ogorman_2013b} and for modulating changes in precipitation over land under global warming \citep{chadwick_et_al_2013, byrne_ogorman_2015}, and they may affect projected increases in heat stress \citep[e.g.,][]{sherwood_huber_2010}. Despite this importance, a clear understanding of what controls land relative humidity is lacking. Here, we introduce a conceptual model based on boundary-layer moisture balance to analyze changes in land relative humidity, and we apply this model to idealized and full-complexity general circulation model (GCM) simulations.

We first review the energy balance argument for the small increase in relative humidity over ocean \citep{held_soden_2000, schneider_2010} and why it does not apply over land. Ocean evaporation is strongly influenced by the degree of sub-saturation of near-surface air, and changes in ocean relative humidity with warming may be estimated from the changes in evaporation using the bulk formula for evaporation, provided that the air-surface temperature disequilibrium and the changes in the exchange coefficient and surface winds are negligible. \citet{schneider_2010} used this approach, together with an energetic estimate for changes in evaporation, to yield an increase in ocean relative humidity with warming of order $1\%\,\rm{K}^{-1}$ (here and throughout this chapter, relative humidity changes are expressed as absolute rather than fractional changes). The simulated increases over ocean are generally smaller (Fig.~\ref{fig:delta_hurs_CMIP5}), indicating that effects such as changes in surface winds must also play a role \citep[e.g.,][]{richter_xie_2008}.

This approach to understanding the increases in ocean relative humidity under warming relies on there being a simple energetic constraint on changes in evaporation, and these evaporation changes being easily related to changes in temperature and surface-air relative humidity. These two conditions are generally not valid over land, where the moisture supply for evapotranspiration is limited and varies greatly across continents \citep{de_jeu_et_al_2008}. The spatially inhomogeneous response of soil moisture to global warming, in addition to changes in land use and stomatal conductance under elevated $\rm{CO}_2$ concentrations \citep[e.g.,][]{sellers_et_al_1996,piao_et_al_2007,cao_et_al_2010,andrews_et_al_2011,cronin_2013}, lead to land evapotranspiration changes with substantial spatial structure \citep{laine_et_al_2014}, and the near-surface relative humidity is merely one of many factors influencing evapotranspiration changes.

To understand the simulated decreases in land relative humidity under global warming, we take a different approach following \citet{rowell_jones_2006}, \citet{simmons_et_al_2010}, \citet{ogorman_muller_2010} and \citet{sherwood_fu_2014}, who discuss how the land boundary-layer specific humidity is influenced by the moisture transport from the ocean. Under global warming, as continents warm more rapidly than oceans, the rate of increase of the moisture transport from ocean to land cannot keep pace with the faster increase in saturation specific humidity over land, implying a drop in land relative humidity. This explanation is attractive because it relies on robust features of the global warming response, namely the small changes in relative humidity over ocean and the stronger surface warming over land. Indeed, the most recent Intergovernmental Panel on Climate Change (IPCC) report cites this argument to explain both observed and projected land relative humidity decreases with warming \citep[][see Section 12.4.5.1 therein]{ipcc_ar5_wg1_chap_12}. However, this explanation has not been investigated quantitatively using either observations or climate models. Thus, it not clear to what extent changes in land relative humidity can be understood as a simple consequence of the land-ocean warming contrast and changes in moisture transport from ocean to land. Indeed, changes in evapotranspiration resulting from soil moisture decreases \citep{berg_et_al_2016} and stomatal closure \citep{cao_et_al_2010} have been shown to influence land relative humidity, though such effects are not considered in the simple argument outlined above.

\begin{figure}[t]
  \centering
\noindent\includegraphics[width=19pc,angle=0]{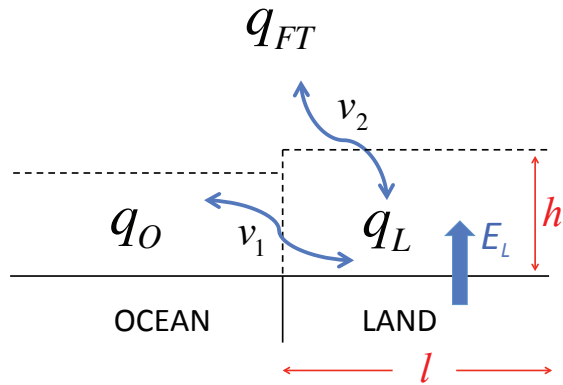}
  \caption{Schematic diagram of the processes involved in the moisture budget of the boundary layer above a land surface [see text and equation (\ref{eqn:box_model_1}) for definitions of the various quantities].}
\label{fig:schematic_theory_1}
\end{figure}

Changes in land surface properties that affect evapotranspiration may affect relative humidity through induced changes in surface-air temperature as well as through changes in specific humidity. Previous studies have shown that soil drying or decreases in stomatal conductance lead to an increase in surface temperature \citep[e.g.,][]{sellers_et_al_1996,seneviratne_et_al_2010,cao_et_al_2010,andrews_et_al_2011, seneviratne_et_al_2013} and this is typically argued to be a result of decreased evaporative cooling of the land surface. But it is difficult to make a quantitative theory for the increase in temperature from the surface energy budget because the surface energy fluxes depend on multiple factors over land, and the effect of increased surface sensible heat flux on surface-air temperature cannot be estimated without taking into account atmospheric processes such as convection. The changes in the land surface-air temperature may be instead related in a straightforward way to changes in surface relative humidity under climate change by using the fact that atmospheric processes constrain the surface-air equivalent potential temperature \citep{byrne_ogorman_2013,byrne_ogorman_2013b}. In particular, changes in surface-air temperature and relative humidity combine to give approximately equal increases in equivalent potential temperature over land and ocean. This link between land and ocean is a result of atmospheric dynamical constraints on vertical and horizontal temperature gradients in the atmosphere [see also \citet{joshi_2008_2}], and it implies that there can be a strong feedback over land between decreases in relative humidity and increases in surface-air temperature. This temperature-relative humidity feedback 
is distinct from soil moisture-temperature or soil moisture-precipitation feedbacks that may also be operating \citep[e.g.,][]{seneviratne_et_al_2010},  and here we assess its importance for decreases in land relative humidity using the atmospheric dynamic constraint discussed above.

We first derive a conceptual box model for the moisture balance of the land boundary layer (section 2). We apply the box model to idealized GCM and Coupled Model Intercomparison Project 5 (CMIP5) simulations, using first a simplified ocean-influence version of the box model (section 3) and then taking into account evapotranspiration (section 4). We then discuss the feedback between temperature and relative humidity changes over land (section 5), before summarizing our results (section 6).


\section{Box model of the boundary-layer moisture balance over land}

The box model is of the moisture balance of the atmospheric boundary layer above land (see schematic, Fig. \ref{fig:schematic_theory_1}). 
The specific humidity of the boundary layer is assumed to be determined by three processes: (i) horizontal mixing with the boundary layer over ocean (e.g., via mean-wind advection, diurnal sea breeze), (ii) vertical mixing with the free troposphere (via large-scale vertical motion, turbulent entrainment, shallow and deep convection), and (iii) evapotranspiration. The time evolution of the land boundary-layer specific humidity, $q_{\rm{L}}$, can then be written as:
\begin{equation}
l h \frac{d q_{\rm{L}}}{dt} = h v_1 (q_{\rm{O}} - q_{\rm{L}}) + l v_2 (q_{\rm{FT}} - q_{\rm{L}}) + \frac{l}{\rho_a}E_{\rm{L}},
\label{eqn:box_model_1}
\end{equation}
where $l$ is the horizontal length scale of the land, $h$ is the depth of the boundary layer over land, $v_1$ and $v_2$ are horizontal and vertical mixing velocities, respectively, $q_{\rm{O}}$ is the specific humidity of the ocean boundary layer, $q_{\rm{FT}}$ is the specific humidity of the free troposphere, $\rho_a$ is the density of air, and $E_{\rm{L}}$ is the evapotranspiration from the land surface. For convenience, we define $\tau_1 = l/v_1$ and $\tau_2 = h/v_2$ as horizontal and vertical mixing timescales, respectively. Lateral advection in the free-troposphere is assumed to lead to equal $q_{\rm{FT}}$ over land and ocean. 
We further assume that the free-tropospheric specific humidity is proportional to the ocean boundary-layer specific humidity, i.e. $q_{\rm{FT}} = \lambda q_{\rm{O}}$, where $\lambda$ is the constant of proportionality. Taking the steady-state solution of (\ref{eqn:box_model_1}) then gives
\begin{equation}
q_{\rm{L}} = \underbrace{\frac{\lambda \tau_1 + \tau_2}{\tau_1 + \tau_2}}_{\gamma} q_{\rm{O}} + \underbrace{\frac{\tau_1 \tau_2}{\rho_a h (\tau_1 + \tau_2)} E_{\rm{L}}}_{q_{\rm{E}}},
\label{eqn:new_box_model_2}
\end{equation}
where we have defined the parameter $\gamma = (\lambda \tau_1 + \tau_2)/(\tau_1 + \tau_2)$
to quantify the influence of ocean specific humidity on land specific humidity, and where
$q_{\rm{E}} =  (\tau_1 \tau_2) E_{\rm{L}} /(\rho_a h (\tau_1 + \tau_2))$ represents the influence of evapotranspiration on land specific humidity.

\begin{figure}[t]
  \centering
  \noindent\includegraphics[width=19pc,angle=0]{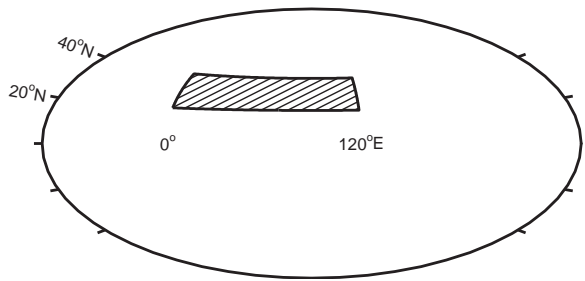}
  \caption{Continental configuration in the idealized GCM simulations. A subtropical continent spans $20^{\circ}\rm{N}$ to $40^{\circ}\rm{N}$ and $0^{\circ}\rm{E}$ to $120^{\circ}\rm{E}$, with a slab ocean elsewhere.}
\label{fig:continent_schematic}
\end{figure}

Other idealized mixed-layer models, generally more complicated than the box model described above, have been previously developed to study land-atmosphere interactions \citep[e.g.,][]{brubaker_entekhabi_1995,betts_2000,joshi_2008_2}. 
Our box model must be taken as a time average over the diurnal cycle in the boundary layer [see discussion in \citet{betts_2000}]. A more complicated model could include the strong diurnal cycle over land, as well as explicitly accounting for the effects of the difference in maximum boundary layer depth over land and ocean \citep[e.g.,][]{von_engeln_teixeira_2013}.

\section{Ocean-influence box model}

For the simplest version of our box model, the ``ocean-influence box model'', we assume that the influence of evapotranspiration on the boundary-layer moisture balance over land is negligible. Setting $E_{\rm{L}}=0$ in (\ref{eqn:new_box_model_2}), we find:
\begin{equation}
q_{\rm{L}} \approx \gamma q_{\rm{O}}.
\label{eqn:box_model_2}
\end{equation}
The parameter $\gamma$ may remain approximately constant under climate change even if there are changes in the mixing time scales $\tau_1$ and $\tau_2$. For example, if the overall tropical circulation and convective mass fluxes slow down with climate warming \citep[e.g.,][]{held_soden_2006,vecchi_soden_2007} such that both mixing time scales increase by the same factor, then this will not cause $\gamma$ to change. Assuming negligible changes in $\gamma$, we can write:
\begin{equation}
\delta q_{\rm{L}} \approx \gamma \delta q_{\rm{O}},
\label{eqn:box_model_3}
\end{equation}
where $\delta$ denotes the change in climate.

The same results would also follow (with a different definition of $\gamma$) if the influence of evapotranspiration on land specific humidity, $q_{\rm{E}}$, is not neglected but is instead assumed to scale with land specific humidity. 
Note that $q_{\rm{E}}$ scaling with land specific humidity is not
the same as the evapotranspiration rate scaling with land specific humidity.
In particular, $q_E$ depends on  
$(\tau_1 \tau_2)/(\tau_1 + \tau_2)$, and this factor would change even if
both $\tau_1$ and $\tau_2$ increase by the same factor.

The ocean-influence box model suggests a straightforward hypothesis - that the ratio of land to ocean specific humidity remains approximately constant as the climate changes or, equivalently, that fractional changes in specific humidity over land and ocean are equal: 
\begin{equation}
\frac{\delta q_{\rm{L}}}{q_{\rm{L}}} \approx \frac{\delta q_{\rm{O}}}{q_{\rm{O}}}.
\label{eqn:constant_q_ratio}
\end{equation}
By contrast the fractional changes in saturation specific humidity depend on the local temperature change and will be bigger over land than ocean.
To the extent that (\ref{eqn:constant_q_ratio}) holds, it is clear that if land warms more than ocean, and ocean relative humidity does not change greatly, then the land relative humidity will decrease.

We now assess the applicability of this ocean-influence box model result to idealized and comprehensive GCM simulations. 
We use (\ref{eqn:box_model_3}) to estimate the change in land relative humidity under climate change given the changes in land temperature
and ocean specific humidity, and calculating $\gamma$ as the the ratio of land to ocean specific humidities in the control climate.

\begin{figure}[t]
  \centering
  \noindent\includegraphics[width=19pc,angle=0]{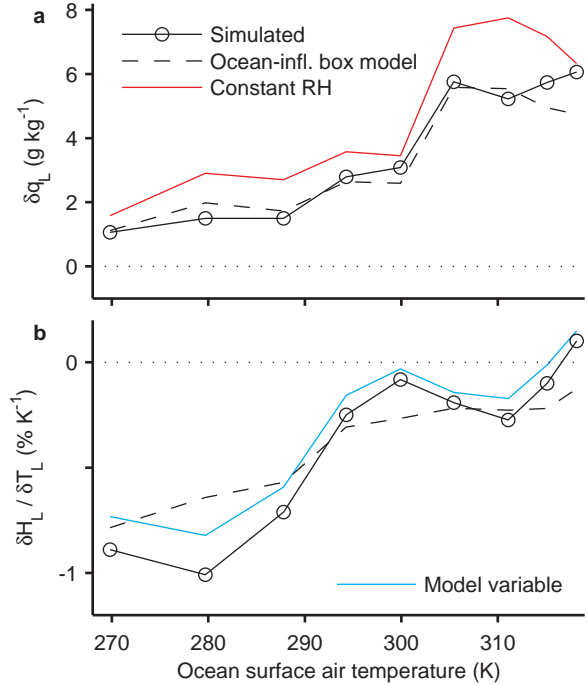}
  \caption{Changes over land in (a) surface-air specific humidity and (b) surface-air relative humidity between pairs of idealized GCM simulations with a subtropical continent.  The relative humidity changes are 
normalized by the land surface-air temperature change.
Solid black lines denote the simulated changes and the dashed lines represent the estimated changes using the ocean-influence box model (\ref{eqn:box_model_3}). Pseudo relative humidities are shown (see text), but the blue line in (b) shows the change in the mean of the actual relative humidity for comparison. The red line in (a) indicates what the change in surface-air land specific humidity would be if the land pseudo relative humidity did not change (i.e. for each pair of simulations, the land specific humidity change if the land pseudo relative humidity is fixed at its value in the colder simulation).}
\label{fig:delta_q_L_and_RH_L_theory_1}
\end{figure}


\subsection{Application of ocean-influence box model to idealized GCM simulations}
\label{sect:iGCM_analysis_theory_1}

The ocean-influence box model is first applied to idealized GCM simulations over a wide range of climates. The idealized GCM is similar to that of \citet{frierson_2006} and \citet{frierson_2007}, with specific details as in \citet{byrne_ogorman_2013} and \citet{ogorman_2008}. It is based on a spectral version of the GFDL dynamical core, with a two-stream gray radiation scheme, no cloud or water vapor radiative feedbacks, and the simplified moist convection scheme of \citet{frierson_2007}. The simulations have a subtropical continent spanning $20^{\circ}\rm{N}$ to $40^{\circ}\rm{N}$ and $0^{\circ}\rm{E}$ to $120^{\circ}\rm{E}$, with a slab ocean elsewhere (Fig. \ref{fig:continent_schematic}). The land surface hydrology is simulated using a simple bucket model \citep{manabe_1969} and all other land surface properties are identical to those of the slab ocean. We vary the climate over a wide range of global-mean surface-air temperatures (between 260K and 317K) by changing the longwave optical thickness, which is analogous to varying the concentrations of $\rm{CO}_2$ and other greenhouse gases. The longwave optical thickness is specified by $\tau = \alpha \tau_{\rm{ref}}$, where $\tau_{\rm{ref}}$ is a reference optical thickness distribution, and we analyze simulations with 10 different values of the parameter $\alpha$\footnote{Simulations are performed with the following $\alpha$ values: 0.2, 0.4, 0.7, 1.0, 1.5, 2.0, 3.0, 4.0, 5.0, and 6.0.}. We present results based on time averages over 4000 days.

When applying the box model to the simulations, we assume that the specific humidity is well-mixed in the boundary layer and use the surface-air specific humidities to represent the entire boundary layer. In the case of the idealized GCM, surface-air quantities are taken to be those of the lowest atmospheric level, $\sigma = 0.989$, where $\sigma=p/p_s$, and $p$ and $p_s$ are the pressure and surface pressure, respectively. The results are qualitatively similar when specific humidities are instead averaged between the surface and $\sigma = 0.9$ (not shown). Land values are averaged (with area weighting) over the entire subtropical continent, and the ocean averages are taken over neighboring ocean at the same latitudes, i.e. from $20^{\circ}\rm{N}$ to $40^{\circ}\rm{N}$ and $120^{\circ}\rm{E}$ to $360^{\circ}\rm{E}$.\footnote{
Our results are almost identical if we calculate ocean averages using the ``control'' Southern Hemisphere as in \citet{byrne_ogorman_2013}, i.e. ocean values averaged over $20^{\circ}\rm{S}$ to $40^{\circ}\rm{S}$ and $0^{\circ}\rm{E}$ to $120^{\circ}\rm{E}$. We choose to average over neighboring ocean in this study because the box model involves advection of moisture from ocean to land, and this naturally suggests averaging over ocean adjacent to the land continent.}

\begin{figure}[t]
  \centering
  \noindent\includegraphics[width=19pc,angle=0]{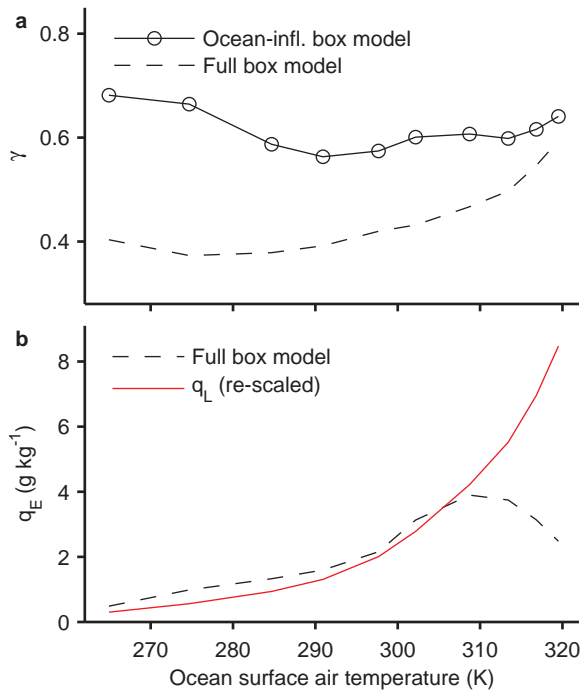}
  \caption{Parameters for the box models applied to the idealized GCM simulations. (a) The $\gamma$ parameter for the ocean-influence box model (solid black line) and the full box model (dashed black line). (b) The $q_{\rm{E }}$ parameter for the full box model (dashed black line) and the surface-air land specific humidity (red line) scaled by a factor of 0.25 so that is roughly matches the magnitude of $q_{\rm{E}}$.}
\label{fig:gamma_and_epsilon}
\end{figure}

To apply the ocean-influence box model (\ref{eqn:box_model_3}), we calculate the $\gamma$ parameter (the ratio of land to ocean specific humidities) for each simulation (except the warmest). We then estimate the change in surface-air land specific humidity between pairs of nearest-neighbor simulations as a function of $\gamma$ and the changes in ocean specific humidity, where $\gamma$ is set to its value in the colder of the two simulations and assumed to be constant as the climate changes. 

Land surface-air specific humidity changes between the pairs of idealized GCM simulations, along with the estimates of these changes using (\ref{eqn:box_model_3}), are plotted against the mid-point ocean temperature for each pair in Fig. \ref{fig:delta_q_L_and_RH_L_theory_1}.  The simulated specific humidity changes are well-captured by the ocean-influence box model (\ref{eqn:box_model_3}) over the full range of climates (Fig. \ref{fig:delta_q_L_and_RH_L_theory_1}a).  
The increases in specific humidity, $\delta q_{\rm{L}}$, generally increase in magnitude as the climate warms (Fig. \ref{fig:delta_q_L_and_RH_L_theory_1}a),
and they are generally smaller than what would occur if land relative humidity remained constant (see the red line in Fig. \ref{fig:delta_q_L_and_RH_L_theory_1}a). 
The small deviations from the prediction of the ocean-influence box model (\ref{eqn:box_model_3}) could be due to the influence of evapotranspiration, changes in circulation patterns, or changes in the ratio $\lambda$ of free tropospheric to surface-air ocean specific humidity which is assumed to be constant in the box model. The parameter $\lambda$ might be expected to increase with warming because the fractional rate of increase in saturation vapor pressure with temperature is higher at the lower temperatures that occur further up in the atmosphere, and
because there is enhanced warming aloft at low latitudes in simulations of global warming \citep[e.g.,][]{santer_et_al_2005},
and such effects could be included in a more complicated box model. 

The $\gamma$ parameter is relatively constant over the wide range of climates simulated (Fig. \ref{fig:gamma_and_epsilon}a) consistent with our neglect of changes in $\gamma$ when deriving (\ref{eqn:box_model_3}), with a mean value of 0.61 and minimum and maximum values of 0.56 and 0.68, respectively. Thus, for the subtropical continent in this idealized GCM, land specific humidity is approximately 60\% of the neighboring ocean specific humidity.

\begin{figure}[t]
  \centering
  \noindent\includegraphics[width=19pc,angle=0]{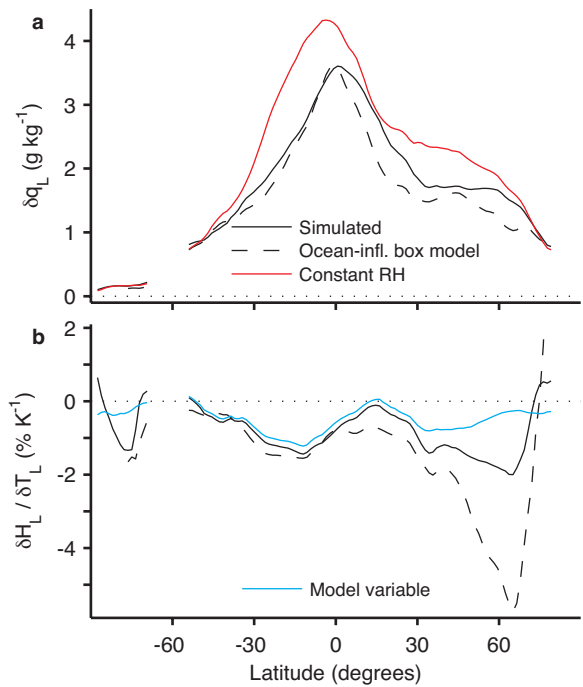}
  \caption{Multimodel-mean changes between the historical and RCP8.5 simulations in zonal and time mean (a) surface-air land specific humidity and (b) surface-air land relative humidity normalized by the global-mean surface-air temperature change. Solid black lines denote the simulated changes and dashed lines denote the estimated changes using the box model (\ref{eqn:box_model_3}). For (a), the red line indicates the change in surface-air land specific humidity for constant land pseudo relative humidity (i.e. for land pseudo relative humidity fixed at the values in the historical simulations). 
Pseudo relative humidities are shown, but the blue line in (b) shows the simulated mean changes for the relative humidity variable outputted by the models for comparison.}
\label{fig:delta_q_RH_CMIP5_sim_theory}
\end{figure}

The box model (\ref{eqn:box_model_3}) predicts the changes in mean specific humidity which must be combined with the mean temperatures to estimate the relative humidity changes. However, because of the nonlinearity of the Clausius-Clapeyron relation, it is not possible to reproduce the mean relative humidity using the mean temperature and mean specific humidity. We instead use a pseudo relative humidity, defined in terms of the mean temperature, specific humidity, and pressure as $H(\overline{T},\overline{p},\overline{q})$ where the bars denote time and spatial means, and $H(T,p,q)$ is the thermodynamic relationship between relative humidity, temperature, pressure and specific humidity. For convenience we will refer to this pseudo relative humidity as the relative humidity, but we also show the actual relative humidity changes for comparison in Figs. \ref{fig:delta_q_L_and_RH_L_theory_1}b and \ref{fig:delta_q_RH_CMIP5_sim_theory}b.

The box-model estimate of the relative humidity changes is somewhat less accurate than for specific humidity, but the decreases in relative humidity with warming and the decreasing magnitude of these changes as the climate warms are both captured (Fig. \ref{fig:delta_q_L_and_RH_L_theory_1}b). 

Given the simplicity of the ocean-influence box model, its ability to describe the behavior of land relative humidity in this idealized GCM is impressive. However, the geometry and surface properties of Earth's land masses are more varied and complex than the idealized continent considered, and factors such as orography or cloud feedbacks that are not included in the idealized GCM could alter the surface humidity response. Therefore, to investigate the changes in land relative humidity further, we turn to more comprehensive simulations from the CMIP5 archive.

\subsection{Application of ocean-influence box model to CMIP5 simulations}
\label{sect:CMIP5_analysis_theory_1}

We apply the ocean-influence box model to changes in land surface-air relative humidity between 30 year time averages in the historical (1976-2005) and RCP8.5 (2070-2099) simulations from the CMIP5 archive \citep{taylor_et_al_bams_2012}. We analyze 19 models in total\footnote{The CMIP5 models considered are: ACCESS1-0, ACCESS1-3, BCC-CSM1-1, BCC-CSM1-1-M, BNU-ESM, CanESM2, CNRM-CM5, CSIRO-Mk3-6-0, GFDL-CM3, GFDL-ESM2M, INMCM4, IPSL-CM5A-LR, IPSL-CM5A-MR, IPSL-CM5B-LR, MIROC-ESM, MIROC-ESM-CHEM, MIROC5, MRI-CGCM3, and NorESM1-M. The variables used in this paper have the following names in the CMIP5 archive: evaporation (\textit{evspsbl}), surface-air specific humidity (\textit{huss}), surface-air temperature (\textit{tas}), and surface-air relative humidity (\textit{hurs}).}, and in each case the r1i1p1 ensemble member is used. As for the idealized GCM analysis, we assume moisture is well mixed in the boundary layer and take surface-air specific humidity to be representative of the boundary layer (using the average specific humidity between the surface and 900hPa gives similar results).

The specific humidities in the box model are identified with the zonal and time mean specific humidities (over land or ocean) for each latitude and for each of the (12) months of the year in the CMIP5 simulations.  We calculate $\gamma$ as the ratio of the mean land and ocean specific humidities at each latitude and for each month of the year in the historical simulations. By computing $\gamma$ in this way, we are assuming that the horizontal exchange of moisture between land and ocean, described by the box model, is taking place predominantly in the zonal direction. 
Using the diagnosed $\gamma$, and assuming it does not change as the climate warms, changes in mean surface-air land specific humidity are estimated for each latitude and each month of the year using (\ref{eqn:box_model_3}) and the changes in mean ocean specific humidity.

\begin{figure}[t]
  \centering
  \noindent\includegraphics[width=19pc,angle=0]{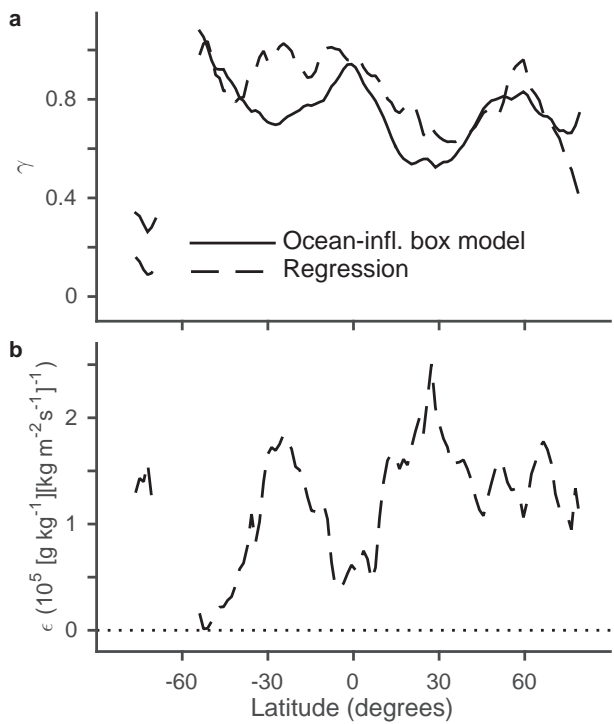}
  \caption{Multimodel means in the CMIP5 simulations of the (a) $\gamma$ parameter for the ocean-influence box model (solid black line) and for the regression approach including evapotranspiration (dashed black line), and (b) the regression coefficient $\epsilon$. The parameter $\gamma$ is evaluated based on the historical simulations for the ocean-influence box model, and $\gamma$ and $\epsilon$ are evaluated using equation \ref{eqn:regression} for the regression approach.}
\label{fig:gamma_eps_zeta_vs_lat}
\end{figure}


The simulated and estimated annual mean changes in land specific humidity at each latitude are shown in Fig. \ref{fig:delta_q_RH_CMIP5_sim_theory}a, and the magnitude and latitudinal variations of the changes are reasonably well captured by the ocean-influence box model, including the flat region in the Northern Hemisphere midlatitudes. The magnitude of the increases is underestimated at most latitudes which, as discussed in the case of the idealized GCM simulations could be partly due to increases in the parameter $\lambda$ relating free-tropospheric specific humidity to ocean surface specific humidity, but other aspects of the ocean-influence box model such as neglecting the influence of evapotranspiration are also likely to play a role. The parameter $\gamma$ (i.e. the ratio of land and ocean specific humidities) is shown in Fig.~\ref{fig:gamma_eps_zeta_vs_lat}a. It has a global, annual and multimodel mean value of 0.74, which is somewhat larger than the
value found in the idealized GCM simulations. This is not
surprising given that the land in the idealized simulations
is a subtropical continent, which is generally drier relative
to neighboring oceans than continents at lower or higher
latitudes. 


Together with the simulated changes in monthly-mean surface-air land temperature, the estimated changes in specific humidity are used to estimate the land pseudo relative humidity changes. As for the idealized GCM analysis, it is necessary to compare pseudo-relative humidities because of the difficulty in converting time and zonal mean specific humidities estimated by the box model to relative humidities. The use of pseudo relative humidities also avoids the complication that different climate models use different saturation vapor pressure formulations. The changes in pseudo relative humidity are calculated for each month of the year before taking the annual mean for both the simulated changes and the changes estimated by the box model. The changes in pseudo relative humidity and model-outputted relative humidity are somewhat similar at lower latitudes but quite different at higher latitudes (cf. blue and black solid lines in Fig. \ref{fig:delta_q_RH_CMIP5_sim_theory}b), where the differing computations of saturation vapor pressure over ice in the various models becomes important and there is larger variability. Nonetheless the pseudo relative humidity is a useful measure of subsaturation, and we will refer to pseudo relative humidity as relative humidity for simplicity.



The simulated changes in (pseudo) land relative humidity are quite well described by the ocean-influence box model in the Southern Hemisphere and at lower latitudes (Fig. \ref{fig:delta_q_RH_CMIP5_sim_theory}b). The estimated and simulated global-mean land relative humidity changes in the various climate models are also correlated, with a correlation coefficient of 0.66. 
Due to the general underestimation of the specific humidity increases by the ocean-influence box model (Fig. \ref{fig:delta_q_RH_CMIP5_sim_theory}a), the relative humidity decreases are overestimated (Fig. \ref{fig:delta_q_RH_CMIP5_sim_theory}b), with a large discrepancy in the mid- to high-latitudes of the Northern Hemisphere. At these latitudes, there is more land than ocean and it is likely that changes in ocean specific humidity have a weak influence on the specific humidity in the interior of large continents, or that meridional moisture transports from ocean at other latitudes become more important. Changes in relative humidity in these inner continental regions are likely to be more strongly influenced by local evapotranspiration changes, and they could be influenced by
shifts in the iceline or changes in soil moisture or vegetation, effects which are not considered in the ocean-influence box model. 




\section{Influence of evapotranspiration}
\label{sect:full_model_RH_land}


The ocean-influence box model captures much (but not all) of the behavior in vastly more complex GCMs. However, the moisture balance of the land boundary layer is also affected by evapotranspiration, and changes in land surface properties, such as soil moisture or stomatal conductance, can affect evapotranspiration in the absence of any changes in the overlying atmosphere.  For example, changes in stomatal conductance under elevated $\rm{CO}_2$ conditions has been shown to reduce both evapotranspiration and land relative humidity without changes in ocean humidity \citep{andrews_et_al_2011}.

\begin{figure}[t]
  \centering
  \noindent\includegraphics[width=19pc,angle=0]{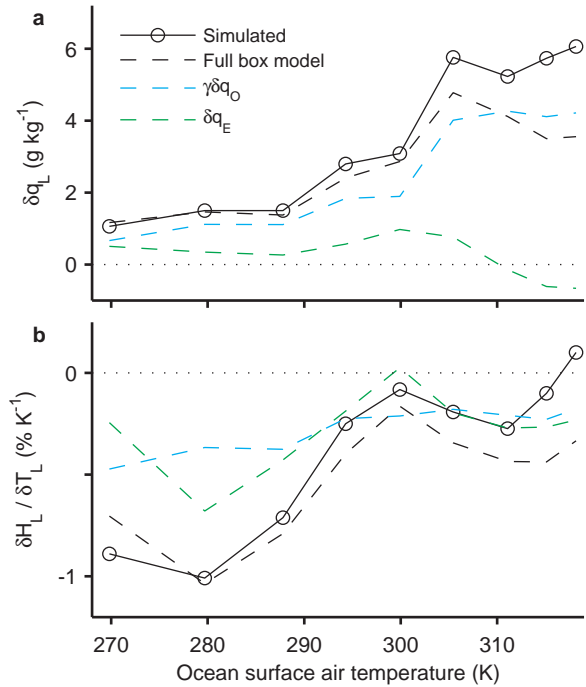}
  \caption{As in Figure \ref{fig:delta_q_L_and_RH_L_theory_1}, but here showing estimates of the surface-air specific and relative humidity changes from the full box model (\ref{eqn:new_box_model_3}). The contributions due to ocean specific humidity changes (blue dashed lines) and evapotranspiration changes (green dashed lines) are also shown. The contributions to changes in relative humidity are calculated using (\ref{eqn:appendix_3}). Pseudo relative humidities are shown in this figure (the changes in actual relative humidity are shown by the blue line in Fig.~\ref{fig:delta_q_L_and_RH_L_theory_1}b.)}
\label{fig:delta_q_L_and_RH_L_theory_2}
\end{figure}




%
%
We turn to the full box model (\ref{eqn:new_box_model_2}) which includes
the effects of evapotranspiration. We assume once more that changes in $\gamma$ are negligible, 
such that:
\begin{equation}
\delta q_{\rm{L}} \approx \gamma \delta q_{\rm{O}} + \delta q_{\rm{E}}.
\label{eqn:new_box_model_3}
\end{equation}
There are two terms contributing to changes in $q_{\rm{L}}$ in (\ref{eqn:new_box_model_3}): the term arising from changes in ocean specific humidity, $\gamma \delta q_{\rm{O}}$, and an additional land evapotranspiration term, $\delta q_{\rm{E}}$. We next assess their relative importance in controlling changes in land humidity in the idealized GCM and CMIP5 simulations.

\subsection{Application of full box model to idealized GCM simulations}
\label{sect:theory_2_simulation_analysis}

We first examine the idealized GCM simulations with a subtropical continent. In contrast to the ocean-influence box model (\ref{eqn:box_model_3}), for which the single parameter $\gamma$ could be easily estimated in the control simulation in each case, the full model (\ref{eqn:new_box_model_3}) has two parameters to be estimated, $\gamma$ and $q_{\rm{E}}$. To estimate these parameters, we perform an additional set of simulations with the same longwave optical thicknesses as in the 10 simulations described previously but with the evapotranspiration set to zero over land. 
Specifying the land evapotranspiration in this way is analogous to drying out the soil. (Note that the change in evapotranspiration affects both the humidity of the atmosphere and the surface energy balance.)
Using these additional simulations with $E_{\rm{L}} = 0$, we can estimate $\gamma$ for each climate using (\ref{eqn:new_box_model_2}): $\gamma = q_{\rm{L},E_{\rm{L}}=0} / q_{\rm{O},E_{\rm{L}}=0}$.  The $\gamma$ values obtained are smaller than those calculated from the control climate for the ocean-influence box model (Fig. \ref{fig:gamma_and_epsilon}a) because the contribution of evapotranspiration to the land specific humidity is now also taken into account.  We then use these $\gamma$ values to estimate $q_{\rm{E}}$ for the original simulations with dynamic land surface hydrology 
: $q_{\rm{E}} = q_{\rm{L}} - \gamma q_{\rm{O}}$. The values of $q_{\rm{E}}$ increase with warming except in hot climates (Fig. \ref{fig:gamma_and_epsilon}b).
Interestingly, the influence of evapotranspiration on land specific humidity as measured by $q_{\rm{E}}$ roughly scales with the land specific humidity except in hot climates (compare the dashed black and solid red lines in Fig. \ref{fig:gamma_and_epsilon}b), and this helps to explain why the ocean-influence box model
is accurate even though evapotranspiration affects land specific humidity.

We then estimate the changes in land specific humidity between pairs of nearest-neighbor simulations from (\ref{eqn:new_box_model_3}) with $\gamma$ assumed to be constant as the climate changes.  The simulated and estimated changes in surface-air land specific humidity, along with the contributions due to changes in ocean specific humidity and land evapotranspiration are shown in Fig. \ref{fig:delta_q_L_and_RH_L_theory_2}a. The full box model captures the basic behavior of the land specific humidity changes as a function of temperature, although it is less accurate in hot climates. The contribution from ocean specific humidity changes, $\gamma \delta q_{\rm{O}}$, is larger than the contribution from land evapotranspiration for all climates (Fig. \ref{fig:delta_q_L_and_RH_L_theory_2}a).  The changes in simulated land (pseudo) relative humidity are also well captured by the full box model (Fig. \ref{fig:delta_q_L_and_RH_L_theory_2}b). 



Because relative humidity depends on temperature as well as specific humidity, 
there is no unique way to use the box model result (\ref{eqn:new_box_model_3}) to decompose changes in land relative humidity into contributions due to ocean specific humidity and land evapotranspiration. However, a decomposition derived in appendix A (equation \ref{eqn:appendix_3}) has several desirable properties. According to the decomposition, the contributions to the change in land relative humidity from evapotranspiration and ocean specific humidity are weighted according to their contribution to land specific humidity in the control climate. The change in ocean specific humidity leads to a decrease in land relative humidity if the fractional increase in ocean specific humidity is less than the fractional increase in saturation specific humidity over land. Similarly, evapotranspiration contributes to a decrease in land relative humidity if the fractional increase in $q_{\rm{E}}$ is less than the fractional increase in saturation specific humidity over land.  (Note that the fractional change in $q_{\rm{E}}$ is generally different from the fractional change in evapotranspiration.)  

Using this decomposition of the change in land relative humidity, we find that the land evapotranspiration contribution is of comparable importance to the ocean specific humidity contribution  for the idealized GCM simulations
(Fig. \ref{fig:delta_q_L_and_RH_L_theory_2}b). By contrast, we found that the
contribution of ocean specific humidity was more important than land evapotranspiration when land specific humidity changes were considered.  The discrepancy arises because, according to the decomposition (\ref{eqn:appendix_3}), it is not the magnitude of a particular contribution to the change in specific humidity that matters for its contribution to the change in relative humidity, but rather how its fractional changes compare to the fractional changes in saturation specific humidity and how much it contributes to the land specific humidity in the control climate. 


\subsection{Influence of evapotranspiration in CMIP5 simulations}

\begin{figure}[t]
  \centering
   \noindent\includegraphics[width=19pc,angle=0]{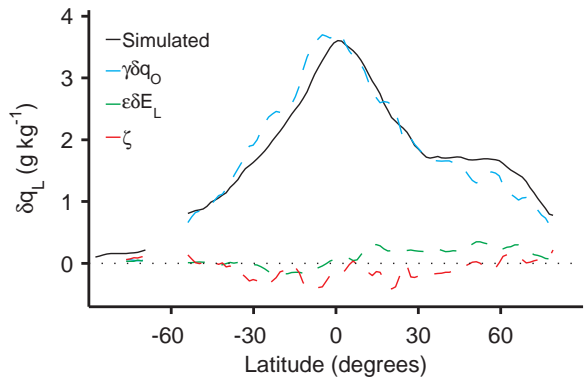}
  \caption{Multimodel-mean changes in zonal- and time-mean surface-air land specific humidity (solid black line) in the CMIP5 simulations, and the contributions to these changes due to ocean specific humidity changes ($\gamma \delta q_{\rm{O}}$; blue dashed line), land evapotranspiration changes ($\epsilon \delta E_{\rm{L}}$; green dashed line), and the remainder term ($\zeta$; red dashed line) as estimated using the regression relation (\ref{eqn:regression}).}
\label{fig:delta_SH_land_CMIP5_theory_2}
\end{figure}

We now investigate how land evapotranspiration contributes to specific humidity changes in the CMIP5 simulations. 
We need to estimate both $\gamma$ and $q_{\rm{E}}$ for the full box model, but there are no CMIP5 simulations analogous to the zero-evapotranspiration simulations with the idealized GCM described above. Instead, we estimate the influence of evapotranspiration and ocean specific humidity on land specific humidity using a multiple linear regression approach based on the intermodel scatter across the CMIP5 models. We use the regression relationship
\begin{equation}
\delta q_{\rm{L}} = \gamma \delta q_{\rm{O}} + \epsilon \delta E_{\rm{L}} + \zeta, 
\label{eqn:regression}
\end{equation}
which is motivated by the full box model (\ref{eqn:new_box_model_2}), but note that $\epsilon \delta E_{\rm{L}}$ will not generally equal $\delta q_{\rm{E}}$ to the extent that parameters such as $\tau_1$ and $\tau_2$ change with climate, and changes in these parameters will contribute to the remainder term $\zeta$.  The variables $\delta q_{\rm{L}},\delta q_{\rm{O}}$ and  $\delta E_{\rm{L}}$ are identified as the zonal and time mean for each latitude and month of the year in each model.  The regression coefficients $\gamma$, $\epsilon$, and $\zeta$ are then estimated using ordinary least squares regression for each latitude and month of the year. The annual means are shown for $\gamma$ and $\epsilon$ in Fig.~\ref{fig:gamma_eps_zeta_vs_lat} and for $\zeta$ in Fig.~\ref{fig:delta_SH_land_CMIP5_theory_2}. The regression coefficient $\gamma$ has a similar magnitude and latitudinal structure to the $\gamma$ parameter calculated for the ocean-influence box model (Fig.~\ref{fig:gamma_eps_zeta_vs_lat}a). The coefficient $\epsilon$ is positive at all latitudes (Fig.~\ref{fig:gamma_eps_zeta_vs_lat}b) indicating that enhanced evapotranspiration increases the land specific humidity, while the remainder term $\zeta$ (Fig.~\ref{fig:delta_SH_land_CMIP5_theory_2}) is negative at most latitudes.

By construction, the regression relationship (\ref{eqn:regression}) is exactly satisfied in the multimodel mean. Based on this relationship, the annual-mean contributions to changes in land specific humidity from changes in ocean specific humidity, changes in land evapotranspiration, and the remainder term are shown in Fig. \ref{fig:delta_SH_land_CMIP5_theory_2}. At all latitudes, changes in land specific humidity are dominated by the ocean specific humidity contribution. The contribution due to changes in land evapotranspiration is positive and has its largest values in the Northern Hemisphere where the land fraction is greatest. 

It is not possible to estimate the individual contributions to changes in land relative humidity for the CMIP5 simulations, as we did for the idealized GCM simulations. This is because the decomposition of relative humidity changes discussed in appendix A involves the individual contributions to land specific humidity in the control climate, and these are difficult to calculate using a regression approach. However, the results from the idealized GCM simulations suggest that evapotranspiration could be important for the changes in land relative humidity in the CMIP5 simulations, even though it is a second order influence for changes in land specific humidity. It would be worthwhile to estimate the land evapotranspiration contribution for full-complexity GCMs by performing simulations with specified land evapotranspiration rates as was done for the idealized GCM in this study.

\section{Feedback between temperature and relative humidity changes over land}

Throughout this paper, we have calculated changes in land relative humidity by first estimating the specific humidity changes and then combining these estimates with the temperature changes, which we have taken as given. However, changes in land relative humidity can be expected to lead to changes in surface-air temperature, and this can be quantified through the atmospheric dynamic constraint linking changes in temperature and relative humidity over land and ocean \citep{byrne_ogorman_2013, byrne_ogorman_2013b}. In the tropics, this constraint is based on convective quasi-equilibrium in the vertical, and weak gradients of free tropospheric temperatures in the horizontal, and extensions to the extratropics are also discussed in \citet{byrne_ogorman_2013b}. As a result, land temperatures and relative humidities must change in tandem as the climate warms such that the change in surface-air equivalent potential temperature ($\theta_e$) is approximately the same over land and ocean ($\delta \theta_{e,\rm{L}} \approx \delta \theta_{e,\rm{O}}$).  Because this constraint follows from atmospheric dynamical processes, we will refer to it as the ``dynamic constraint'' on surface-air temperatures and humidities.
By contrast, we will refer to the link between surface-air humidities over land and ocean because of moisture transport between them (as formulated in the box model in this paper) as the ``moisture constraint'' 
($\delta q_{\rm{L}} = \gamma \delta q_{\rm{O}} + \delta q_{\rm{E}}$).

\begin{figure}[t]
  \centering
  \noindent\includegraphics[trim=0.0in 0.0in 0.0in 0.0in,clip,width=19pc,angle=0]{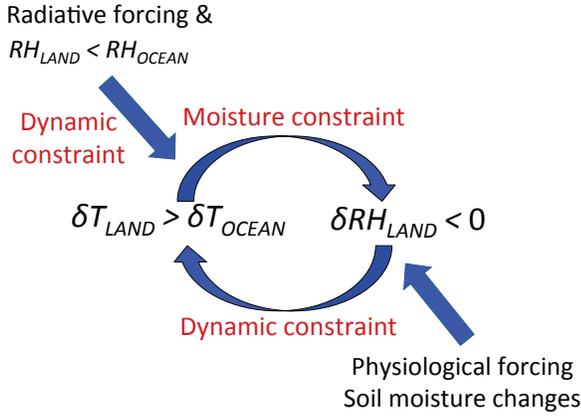}
  \caption{Schematic diagram describing the feedback between changes in temperature and relative humidity over land and ocean (assuming, for simplicity, that ocean relative humidity remains constant). The ``dynamic constraint'' arises from atmospheric processes that link temperatures and relative humidities over land and ocean. The ``moisture constraint'' is due to the limited supply of moisture from the ocean to the land boundary layer.}
\label{fig:RH_T_feedback}
\end{figure}

A feedback loop is used to conceptualize the interaction between changes in temperature and relative humidity over land and ocean (Fig. \ref{fig:RH_T_feedback}). 
Land is drier than ocean in the control climate, and as a result the dynamic constraint implies that land temperatures increase more than the ocean temperature in response to a positive radiative forcing \citep{byrne_ogorman_2013}. The moisture constraint then implies that the enhanced land warming leads to a land relative humidity decrease because of the limited supply of moisture from the ocean. According to the dynamic constraint, a decrease in land relative humidity enhances the land warming further. The feedback loop can also be entered via a non-radiative forcing that causes a decrease in humidity, such as the physiological forcing from reduced stomatal conductance in an elevated-$\rm{CO}_2$ world or a local decrease in soil moisture. 

We assess the strength of the feedback between relative humidity and temperature changes over land for the case in which a forcing alters the specific humidity over land, while the ocean temperature and ocean specific humidity are assumed to remain constant. Considering the relative humidity to be a function of specific humidity and temperature and linearizing, we can write the total change in land relative humidity as the sum of contributions from changes in specific humidity at constant temperature (the ``forced'' component) and changes in temperature at constant specific humidity (the ``temperature feedback'' component):
\begin{equation}
\delta H_{\rm{L},total} = \delta H_{\rm{L},forced} + \left.
\frac{\partial H_{\rm{L}}} {\partial T_{\rm{L}}} \right|_{q_{\rm{L}}} \delta T_{\rm{L}},
\label{eqn:feedback_loop_1}
\end{equation}
where $\partial H_{\rm{L}}/\partial T_{\rm{L}}|_{q_{\rm{L}}}$ is the sensitivity of relative humidity to warming at constant specific humidity, and $\delta T_{\rm{L}}$ is the change in land temperature that arises because of the change in land relative humidity. The land temperature change is then related to the relative humidity change by the dynamic constraint:
\begin{equation}
\delta T_{\rm{L}} = \left. \frac{\partial T_{\rm{L}}} {\partial H_{\rm{L}}} \right|_{\theta_{e,\rm{L}}} \delta H_{\rm{L},total},
\label{eqn:feedback_loop_2}
\end{equation}
where $\partial T_{\rm{L}}/\partial H_{\rm{L}}|_{\theta_{e,\rm{L}}}$ is the sensitivity of temperature to changes in relative humidity at constant equivalent potential temperature ($\theta_{e,\rm{L}}$). The land equivalent potential temperature remains constant because the dynamic constraint requires that changes in equivalent potential temperature are the same over land and ocean [see \citet{byrne_ogorman_2013b}], and we are assuming that the ocean temperatures and humidities (and therefore ocean equivalent potential temperature) are not changing in this example. Combining (\ref{eqn:feedback_loop_1}) and (\ref{eqn:feedback_loop_2}), we can express the total land relative humidity change as:
\begin{equation}
 \delta H_{\rm{L},total} =  \frac{\delta H_{\rm{L},forced}}{1 - \left. \frac{\partial H_{\rm{L}}} {\partial T_{\rm{L}}} \right|_{q_{\rm{L}}} \left. \frac{\partial T_{\rm{L}}} {\partial H_{\rm{L}}} \right|_{\theta_{e,\rm{L}}}}.
\label{eqn:total_RH_change}
\end{equation}

We next evaluate (\ref{eqn:total_RH_change}) for a simple numerical example. For constant specific humidity, a 1K temperature increase leads to approximately a $6\%$ \textit{fractional} reduction in relative humidity. Assuming a land relative humidity of $50\%$ this corresponds to a $3\%$ \textit{absolute} reduction in relative humidity: $\partial H_{\rm{L}} / {\partial T_{\rm{L}}}|_{q_{\rm{L}}} \approx -3\,\%\,\rm{K}^{-1}$. The sensitivity of land temperature to changes in land relative humidity at fixed equivalent potential temperature, $\partial T_{\rm{L}}/\partial H_{\rm{L}}|_{\theta_{e,\rm{L}}}$, can be estimated using the thermodynamic relation $T(\theta_e, H, p)$. For a land relative humidity of $50\%$, a land surface temperature of $298\,\rm{K}$, and a surface pressure of 1000hPa, we find $\partial T_{\rm{L}}/\partial H_{\rm{L}}|_{\theta_{e,\rm{L}}} \approx -0.2 \, \rm{K}\,\%^{-1}$ where we use the \citet{bolton_1980} formulation of $\theta_e$ [note that $\partial T_{\rm{L}} / \partial H_{\rm{L}}|_{\theta_{e,\rm{L}}}$ is plotted as a function of ocean surface temperature in Fig.3a of \citet{byrne_ogorman_2013}]. 

Using (\ref{eqn:total_RH_change}) and the values for the partial derivatives given above, we find that a forced decrease in land relative humidity of $1\%$ results in a total land relative humidity decrease of $2.5\%$ when the temperature feedback is included for this illustrative example, implying an amplification by a factor of 2.5 in the decrease in relative humidity. (For a higher control relative humidities of 70\% over land, the amplification would be by a factor of 3.) Thus, feedbacks between temperature and relative humidity over land strongly amplify forced changes in relative humidity due to, say, changes in stomatal conductance or changes in soil moisture. A further implication is that more than half of the total change in relative humidity comes from the change in temperature rather than the changes in specific humidity, and this holds true for land surface temperatures above 292K in this illustrative example with control land relative humidity of 50\%.


\section{Conclusions}

We have introduced a conceptual box model to investigate the response of near-surface land relative humidity to changes in climate. Neglecting the contribution of evapotranspiration to the moisture balance over land (or assuming that $q_{\rm{E}}$ scales with land specific humidity), the simplest version of our box model suggests a purely oceanic control on land boundary-layer humidity, with equal fractional changes in specific humidity over land and ocean. Together with enhanced warming over land relative to ocean and small changes in ocean relative humidity, this simple box model implies a decrease in land relative humidity as the climate warms. The ocean-influence box model captures many features of the humidity response in idealized GCM and CMIP5 simulations, supporting the hypothesis of a strong oceanic influence on boundary-layer specific humidity over land.

The full box model, incorporating evapotranspiration, is applied to the idealized GCM simulations using additional simulations with specified evapotranspiration rates, and to the CMIP5 simulations using a linear regression approach. Compared to moisture transport from the ocean, evapotranspiration has only a secondary influence on the land specific humidity and its changes.  However, 
according to a decomposition of the relative humidity change that weights the different contributions according to their importance in the control climate, 
evapotranspiration plays an important role for the changes in land relative humidity in the idealized GCM simulations.

In addition, feedbacks between temperature and relative humidity changes over land, associated with the dynamic constraint on the land-ocean warming contrast and the moisture constraint described in this paper, can strongly amplify relative humidity changes. We have derived an expression for the strength of this amplification, and we have given a simple example in which the relative humidity change resulting from a moisture forcing is amplified by a factor of $2.5$ when changes
in temperature are taken into account. For sufficiently high control-climate temperatures, the majority of the change in land relative humidity comes from the induced change in temperature rather than the change in specific humidity.  This amplification is consistent with the strong influence of reduced stomatal conductance or decreases in soil moisture on land relative humidity found in previous studies \citep[e.g.,][]{cao_et_al_2010,andrews_et_al_2011,berg_et_al_2016}.

As mentioned in section 1, the pattern of relative humidity changes influences the response of the water cycle to climate change. In particular, spatial gradients of fractional changes in surface-air specific humidity ($\delta q/q$) contribute a negative tendency to precipitation minus evapotranspiration ($P-E$) over continents as the climate warms \citep{byrne_ogorman_2015}. The ocean-influence box model predicts that $\delta q/q$ is spatially uniform, implying no effect of spatial gradients in this quantity on $P-E$ changes over land. However, the CMIP5 simulations do show spatial gradients in $\delta q/q$, and thus a more detailed understanding of the pattern of relative humidity changes is needed for the purpose of understanding changes in $P-E$ changes over land.
 
Future work could investigate the controls on the detailed pattern of $\delta q/q$ over land in order to better understand the $P-E$ response over land.  
Further investigation of the temperature-relative humidity feedback over land in comprehensive models would also be valuable. Finally, it is of interest to determine if the box models discussed here can be adapted for application to shorter-term variability, and in particular to the sharp decrease in global-mean land relative humidity that is seen in observations between 2000 and 2008 \citep{simmons_et_al_2010,willett_et_al_2014}.

%
\acknowledgments
We thank Alexis Berg, Bill Boos, Jack Scheff, Sonia Seneviratne, and Bjorn Stevens for helpful discussions. We acknowledge the World Climate Research Programme's Working Group on Coupled Modelling, which
is responsible for CMIP, and we thank the climate modeling groups for
producing and making available their model output. For CMIP, the U.S.
Department of Energy's Program for Climate Model Diagnosis and Intercomparison provides coordinating support and led development of software
infrastructure in partnership with the Global Organization for Earth System
Science Portals. We acknowledge support from NSF grant AGS-1148594.

%
\appendix[A]

\appendixtitle{Decomposition of changes in land relative humidity}

In this appendix we derive a decomposition of the changes in land relative humidity into contributions associated with changes in ocean specific humidity and with land evapotranspiration. 

We approximate relative humidity as the ratio of specific humidity to saturation specific humidity, $H \approx q/q^*$, where $q^*$ is the saturation specific humidity. We can then write changes in specific humidity as:

\begin{equation}
\delta q = H \delta q^* + q^* \delta H + \delta H \delta q^*.
\label{eqn:appendix_1}
\end{equation}
Dividing (\ref{eqn:appendix_1}) by the specific humidity $q$ and rearranging, we can express fractional changes in relative humidity as:

\begin{equation}
\frac{\delta H}{H} = \left( \frac{\delta q}{q} - \frac{\delta q^*}{q^*} \right) \frac{q^*}{q^* + \delta q^*}.
\label{eqn:appendix_2}
\end{equation}
Using (\ref{eqn:new_box_model_3}), we relate changes in land specific humidity to changes in ocean specific humidity and changes in evapotranspiration, i.e. $\delta q_{\rm{L}} \approx \gamma \delta q_{\rm{O}} + \delta q_{E}$, and substitute in to (\ref{eqn:appendix_2}) to obtain an expression for fractional changes in land relative humidity in terms of an ocean specific humidity contribution and a evapotranspiration contribution:

\begin{equation}
\begin{split}
\frac{\delta H_{\rm{L}}}{H_{\rm{L}}} &\approx \overbrace{\frac{\gamma q_{\rm{O}}}{q_{\rm{L}}}  \left( \frac{\delta q_{\rm{O}}}{q_{\rm{O}}} - \frac{\delta q_{\rm{L}}^*}{q_{\rm{L}}^*} \right) \frac{q_{\rm{L}}^*}{q_{\rm{L}}^* + \delta q_{\rm{L}}^*}}^{\text{$\gamma \delta q_{\rm{O}}$ contribution}} \\
&+ \underbrace{\frac{q_{\rm{E}}}{q_{\rm{L}}}  \left( \frac{\delta q_{\rm{E}}}{q_{\rm{E}}} - \frac{\delta q_{\rm{L}}^*}{q_{\rm{L}}^*} \right) \frac{q_{\rm{L}}^*}{q_{\rm{L}}^* + \delta q_{\rm{L}}^*}}_{\text{$\delta q_{\rm{E}}$ contribution}}.
\label{eqn:appendix_3}
\end{split}
\end{equation}
%

The properties of this decomposition are discussed in section \ref{sect:full_model_RH_land}\ref{sect:theory_2_simulation_analysis}.

\bibliographystyle{ametsoc2014}
\bibliography{references}

%

%

\end{document}